# On Sexual Selection

Larry Bull

Department of Computer Science & Creative Technologies

University of the West of England, Bristol BS16 1QY, UK

Larry.Bull@uwe.ac.uk

**Abstract**

Sexual selection is a fundamental aspect of evolution for all eukaryotic organisms with mating types. This paper suggests intersexual selection is best viewed as a mechanism to compensate for the unavoidable dynamics of coevolution between sexes that emerge with isogamy. Using the NK model of fitness landscapes, the conditions under which allosomes emerge are first explored. This extends previous work on the evolution of sex where the fitness landscape smoothing of a rudimentary form of the Baldwin effect is suggested as the underlying cause. The NKCS model of coevolution is then used to show how varying fitness landscape size, ruggedness, and connectedness can vary the conditions under which a very simple sexual selection mechanism proves beneficial. This is found to be the case whether one or both sexes exploit sexual selection.

**Introduction**

Sexual selection is the component of natural selection usually referring to both opposite-sex mate choice and same-sex competition for mating opportunities. In the former case, males and females can be seen to be *coevolving* within their species. At an abstract level coevolution is typically considered as the coupling together of the fitness landscapes of the interacting species. Hence the adaptive moves made by one species in its fitness landscape causes deformations in the fitness landscapes of its coupled partner(s). That sexes coevolve and that this need not be to their mutual benefit has been used to explain phenomena such as female-damaging genitalia and sperm selection (eg, [18]). It is beyond the scope of this paper to review the (considerable) literature on sexual selection and the reader is referred to [2] for a recent overview. In this paper Kauffman and Johnsen's [15] NKCS model is used to explore the coevolutionary behaviour of two sexes and how this is affected by mate choice, ie, intersexual selection.

It has been suggested [7] that the emergence of sex – defined as successive rounds of syngamy and meiosis in a haploid-diploid lifecycle - enabled the exploitation of a rudimentary form of the Baldwin effect [4]. Key to this explanation for the evolution of sex in eukaryotes is to view the process from the perspective of the constituent haploids. A diploid organism may been seen to simultaneously represent two points in the underlying haploid fitness landscape. The fitness associated with those two haploids is therefore the fitness achieved in their combined form as a diploid; each haploid genome will have the same fitness value and that will almost certainly differ from that of their corresponding haploid organism due to the interactions between the two genomes. That is, the effects of haploid genome combination into a diploid can be seen as a simple form of phenotypic plasticity for the individual haploids before they revert to a solitary state during reproduction. In this way evolution can be seen to be assigning a single fitness value to the region of the landscape between the two points represented by a diploid's constituent haploid genomes, ie, a simple form of generalization, and hence altering the shape of the haploid fitness landscape. In particular, the latter enables the landscape to be smoothed under the Baldwin effect (after [12]). This paper begins by extending the new view of eukaryotic sex to consider the emergence of allosomes before their subsequent coevolution in mating types.

# The NK Model: Asexual Haploid Evolution

Kauffman and Levin [14] introduced the NK model to allow the systematic study of various aspects of fitness landscapes (see [13] for an overview). In the standard model, the features of the fitness landscapes are specified by two parameters: $N$, the length of the genome; and $K$, the number of genes that has an effect on the fitness contribution of each (binary) gene. Thus increasing $K$ with respect to $N$ increases the epistatic linkage, increasing the ruggedness of the fitness landscape. The increase in epistasis increases the number of optima, increases the steepness of their sides, and decreases their correlation. The model assumes all intragenome interactions are so complex that it is only appropriate to assign random values to their effects on fitness. Therefore for each of the possible $K$ interactions a table of $2^{(K+1)}$ fitnesses is created for each gene with all entries in the range 0.0 to 1.0, such that there is one fitness for each combination of traits (Figure 1). The fitness contribution of each gene is found from its table. These fitnesses are then summed and normalized by $N$ to give the selective fitness of the total genome.

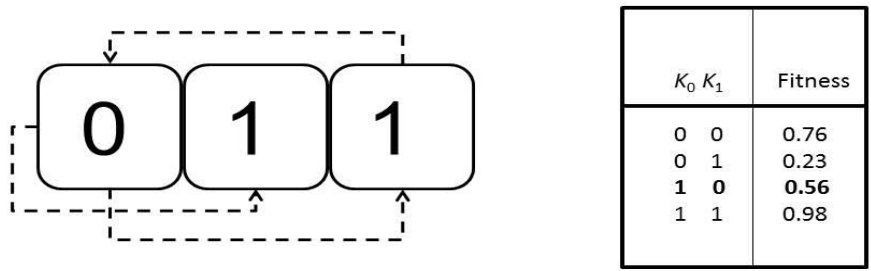

Figure 1. An example NK model ($N=3$, $K=1$) showing how the fitness contribution of each gene depends on $K$ random genes (left). Therefore there are $2^{(K+1)}$ possible allele combinations per gene, each of which is assigned a random fitness. Each gene of the genome has such a table created for it (right, centre gene shown). Total fitness is the normalized sum of these values.

Kauffman [13] used a mutation-based hill-climbing algorithm, where the single point in the fitness space is said to represent a converged species, to examine the properties and evolutionary dynamics of the NK model. That is, the population is of size one and a species evolves by making a random change to one randomly chosen gene per generation. The "population" is said to move to the genetic configuration of the mutated individual if its fitness is greater than the fitness of the current individual; the rate of supply of mutants is seen as slow compared to the actions of selection. Ties are broken at random. Figure 2 shows example results. All NK model results reported in this paper are the average of 10 runs (random start points) on each of 10 NK functions, ie, 100 runs, for 20,000 generations. Here $0 \leq K \leq 15$, for $N=20$ and $N=100$.

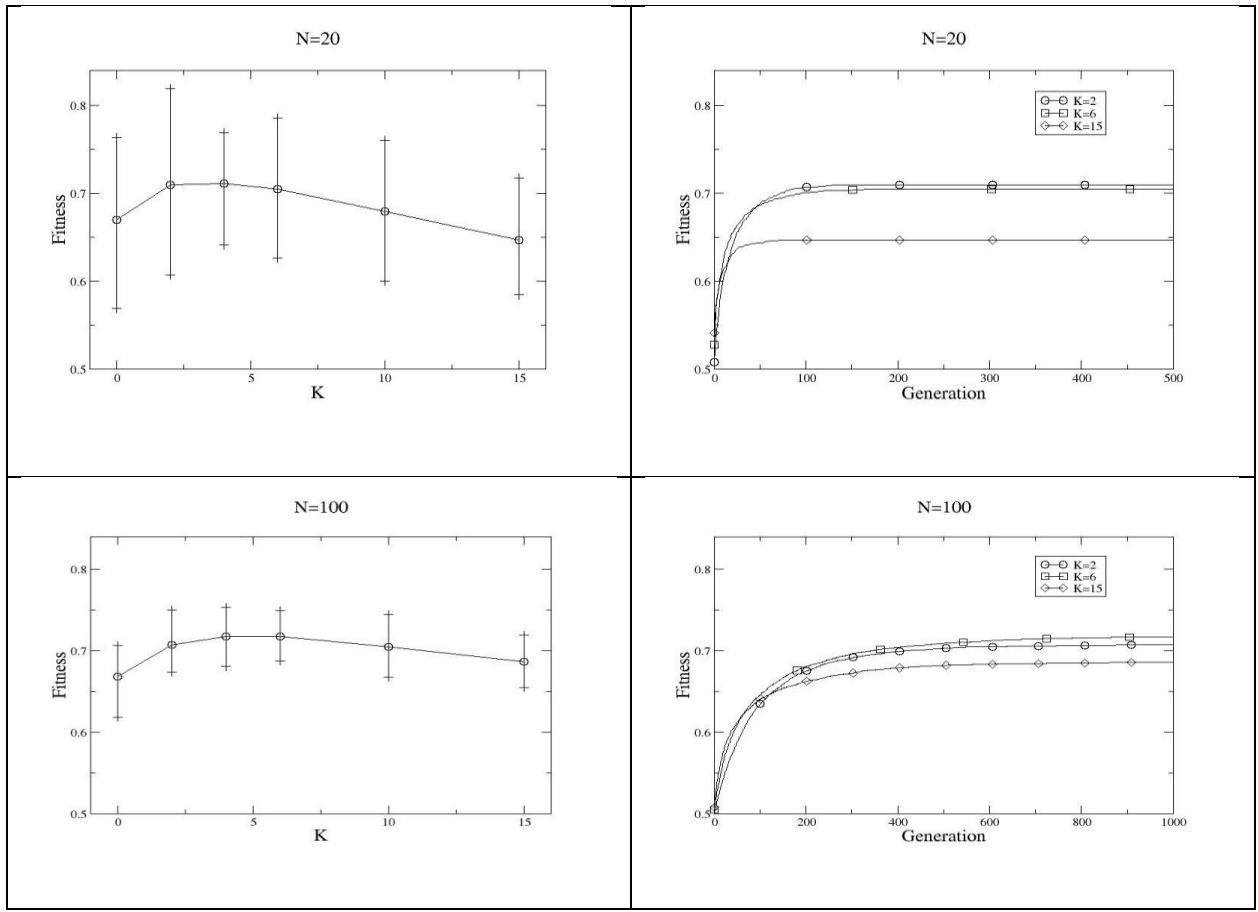

Figure 2. Showing typical behaviour and the fitness reached after 20,000 generations on landscapes of varying ruggedness (*K*) and length (*N*). Error bars show min and max values.

Figure 2 shows examples of the general properties of adaptation on such rugged fitness landscapes identified by Kauffman (eg, [13]), including a "complexity catastrophe" as $K \rightarrow N$. When $K=0$ all genes make an independent contribution to the overall fitness and, since fitness values are drawn at random between 0.0 and 1.0, order statistics show the average value of the fit allele should be 0.66. Hence a single, global optimum exists in the landscape of fitness 0.66, regardless of the value of $N$. At low levels of $K$ ($0<K<8$), the landscape buckles up and becomes more rugged, with an increasing number of peaks at higher fitness levels, regardless of $N$. Thereafter the increasing complexity of constraints between genes means the height of peaks typically found begin to fall as $K$ increases relative to $N$: for large $N$, the central limit theorem suggests reachable optima will have a mean fitness of 0.5 as $K \rightarrow N$. Figure 2 shows how the optima found when $K>6$ are significantly lower for $N=20$ compared to those for $N=100$ (T-test, $p<0.05$).

**Eukaryotes: Sexual Diploid Evolution (Typically)**

Bacteria and archaea can be described as "simpler" than eukaryotes for a variety of reasons including their typically containing a single genome. Hence, as in the NK model above, a given combination of gene values in their genome can be viewed as representing a single point in an $N$-dimensional fitness landscape. In contrast, eukaryotes can contain two or more genomes, typically two, and reproduce sexually via a haploid-diploid cycle with meiosis. Eukaryotes can therefore be viewed as a single point in a fitness landscape of all possible diploids. However, it has recently been suggested that diploid eukaryotes should be viewed as simultaneously representing two points in the fitness landscape of their constituent haploid genomes to explain why sex emerged [7]. Significantly, since the phenotype and hence fitness of a diploid individual is a composite of its two haploid genomes, evolution can be seen to be assigning a single fitness value to the *region* of the landscape between the two points represented by its constituent haploid genomes. That is, it is proposed that evolution is using a more sophisticated scheme by which to navigate the fitness landscapes of eukaryotes than for bacteria: an individual organism provides a generalization in the fitness landscape as opposed to information about a single point. This is seen as particularly useful as the complexity of the fitness landscape increases in both size and ruggedness.

Moreover, the haploid-diploid cycle can also be explained as creating a simple form of the Baldwin effect [4], thereby enabling the beneficial smoothing of rugged fitness landscapes [12]. Briefly, under the Baldwin effect, the existence of phenotypic plasticity potentially enables an organism to display a different (better) fitness than its genome directly represents. Typically, such plasticity is assumed to come from the organism itself, eg, through the modification of neural connectivity. However, a genetically defined phenotype can also be altered by the exploitation of something in the environment, eg, a tool. It is suggested that haploids forming pairs to become diploid is akin to the latter case of learning/plasticity with the partner's genome providing the variation. Being diploid can potentially alter gene expression in comparison to being haploid and hence affect the expected phenotype of each constituent haploid alone since both genomes are active in the cell - through changes in gene product concentrations, partial or co-dominance, etc. (see [17] for related discussions). If the cell/organism subsequently remains diploid and reproduces asexually, there is no scope for a rudimentary Baldwin effect. However, if there is a reversion to a haploid state for reproduction under a haploid-diploid cycle, there is the potential for a significant mismatch between the utility of the haploids passed on compared to that of the diploid selected; individual haploid gametes do not contain all of the genetic material through which their fitness was determined.

The variation processes can then be seen to change the bounds for sampling combined genomes within the diploid by altering the distance between the two end points in the underlying haploid fitness landscape. That is, the degree of possible change in the distance between the two haploid genomes controls the amount of learning possible per cycle. Numerous explanations exist for the benefits of recombination (eg, [5]) but the role becomes clear under the new view: recombination moves the current end points in the underlying haploid fitness space which define the fitness level generalization either closer together or further apart. That is, recombination adjusts the size of an area assigned a single fitness value, potentially enabling higher fitness regions to be more accurately identified over time. Moreover, recombination can also be seen to facilitate genetic assimilation within the simple form of the Baldwin effect: if the haploid pairing is beneficial and the diploid cell/organism is chosen under selection to reproduce, the recombination process can bring an assortment of those partnered genes together into new haploid genomes. In this way the fitter allele values from the pair of partnered haploids may come to exist within individual haploids more quickly than the under mutation alone (see

[7] for full details). Mutation can also be seen to be adjusting the distance between the two genomes at a generally reduced rate per generation. It has previously been shown how both the amount of learning per step and the rate at which it occurs can affect utility, with more learning typically proving increasingly beneficial with increasing $K$ [6].

**Sex in the NK Model**

As discussed in [16, p150] the first step in the evolution of eukaryotic sex was the emergence of a haploid-diploid cycle, probably via endomitosis, then simple syngamy or one-step meiosis, before two-meiosis with recombination. Following [7], the NK model can be extended to consider aspects of the evolution of sexual diploids. Firstly, each individual contains two haploid genomes of the form described above for the standard model. The fitness of an individual is here simply assigned as the average of the fitness of each of its constituent haploids. These are initially created at random, as before. Secondly, simple syngamy is here implemented as follows: on each generation the diploid individual representing the converged population is copied twice to create two offspring. One of the two haploids in each offspring individual is chosen at random. Finally, a random gene in each chosen haploid is mutated. The resulting pair of haploids forms the new diploid offspring to be evaluated.

Two-step meiosis with recombination is here implemented as follows: on each generation the diploid individual representing the converged population is similarly copied twice to create two offspring. In each offspring, each haploid genome is copied once, a single recombination point is chosen at random, and non-sister haploids are recombined. One of the four resulting haploids in each offspring individual is chosen at random. Finally, a random gene in each chosen haploid is mutated. The resulting pair of haploids forms the new diploid offspring to be evaluated (Figure 3).

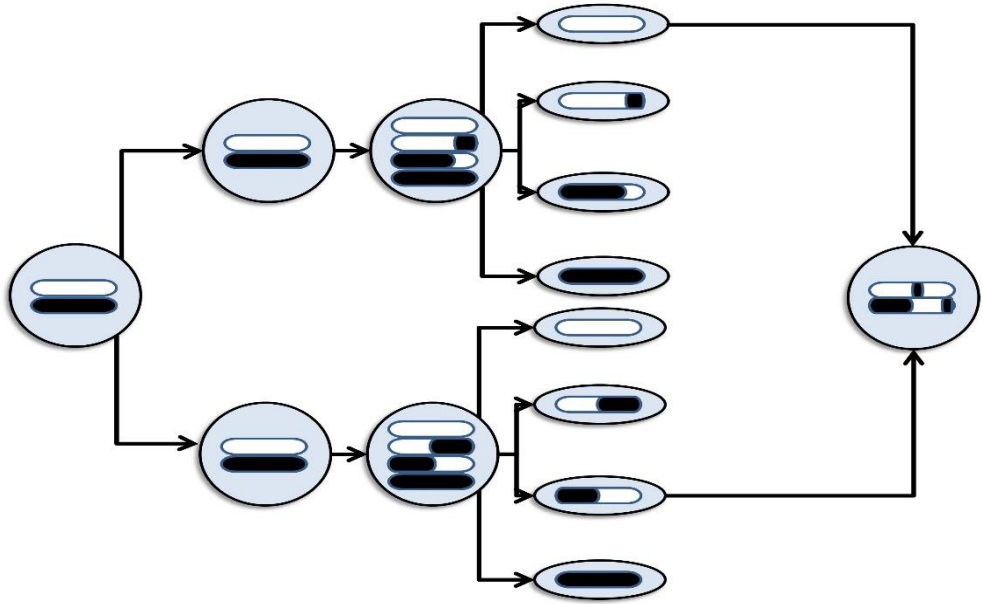

Figure 3. Showing the haploid-diploid cycle with two-step meiosis as implemented with a converged population of hermaphrodites.

Figure 4 shows examples of the general properties of adaptation in the NK model of rugged fitness landscapes for diploid organisms evolving via either simple one-step meiosis or two-step meiosis with recombination. Under both forms of meiosis, when $N=20$ the fitness level reached is significantly lower than for $N=100$ for $K>4$ (T-test, $p<0.05$), as was seen in the traditional haploid case due to the effects of the increased landscape complexity. It can be seen that fitness levels are always higher than the equivalent haploid case (Figure 2) when $K>0$ due to the Baldwin effect as discussed (T-test, $p<0.05$). Again, as reported in [7], the fitness levels are always higher for two-step meiosis with recombination in comparison to simple syngamy for $K>2$ (T-test, $p<0.05$). This is explained by the potential for an increased rate of change in the distance between the two genomes and hence the effective amount of learning experienced per generation (after [6]), as well as the role of assimilation. Figure 5 shows examples of how the size of those generalizations changes over time depending upon the ruggedness of the landscape and the form of meiosis; assimilation through recombination can be seen.

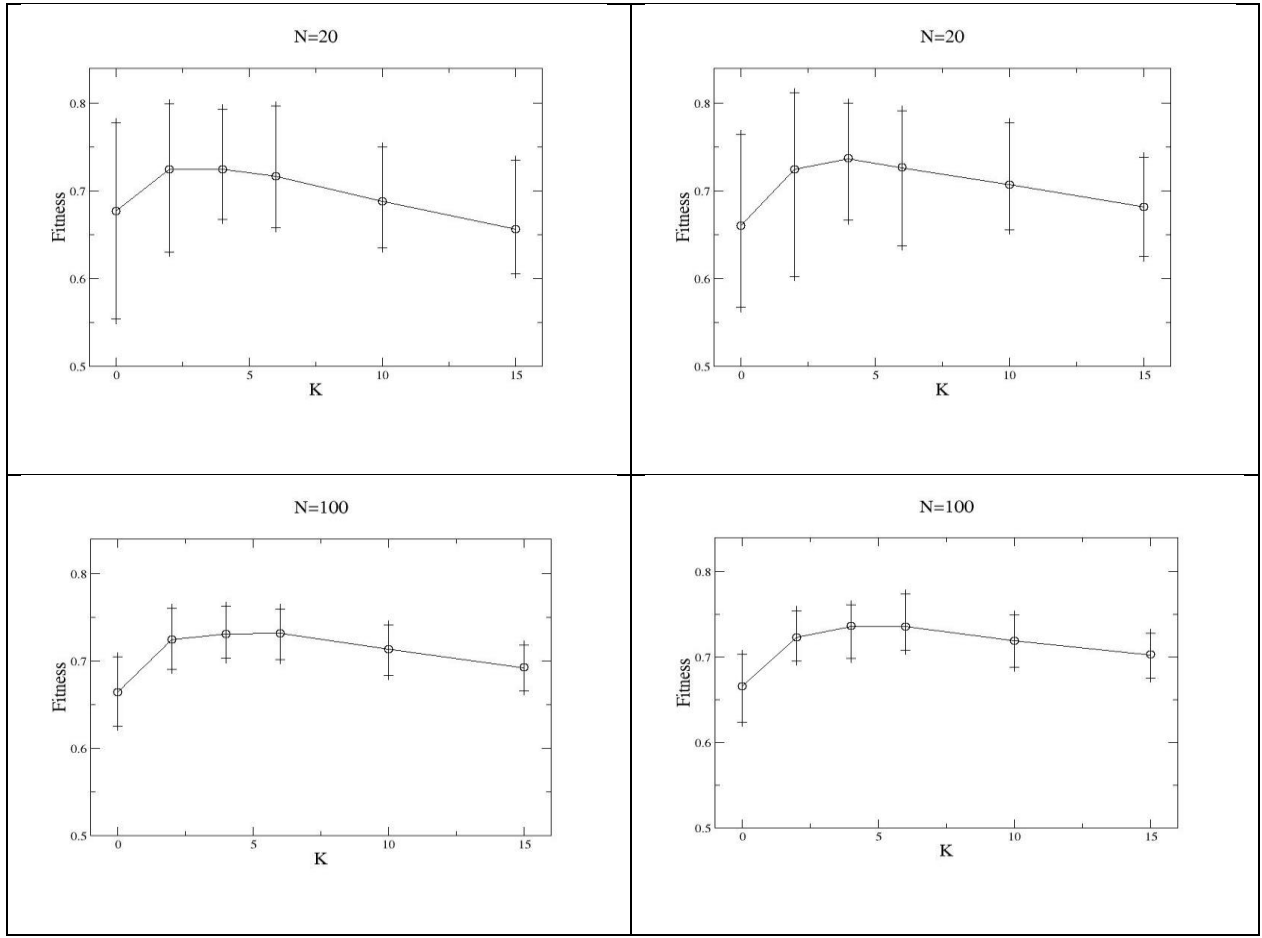

Figure 4. Showing typical behaviour and fitness reached after 20,000 generations on landscapes of varying ruggedness ($K$) and size ($N$) for diploids undergoing one-step meiosis (left) or two-step meiosis (right).

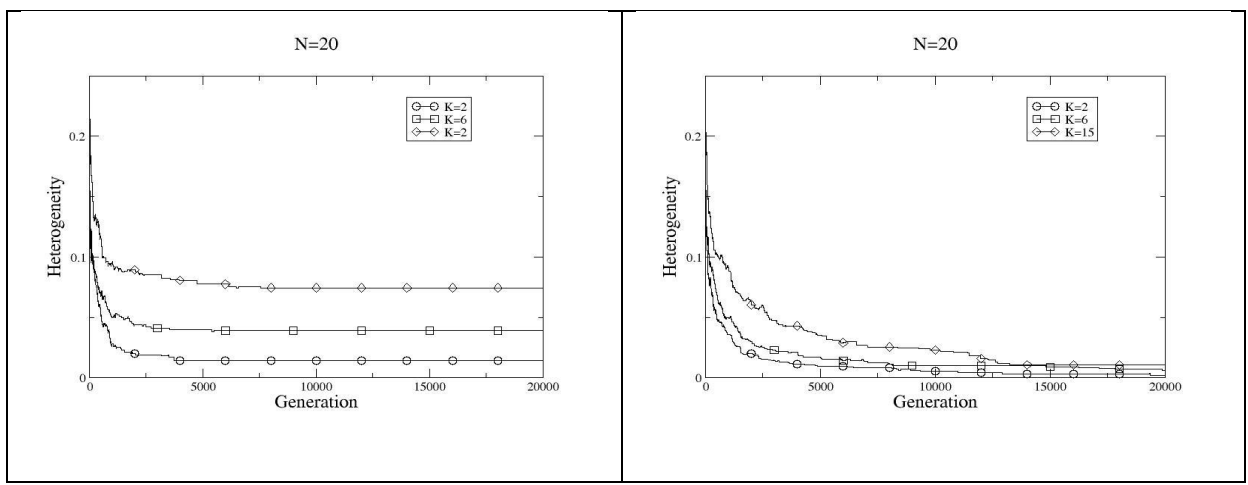

Figure 5. Showing typical convergence behaviour as a fraction of the difference in corresponding gene values in the two haploid genomes on landscapes of varying ruggedness ($K$) for diploids undergoing one-step meiosis (left) and two-step meiosis with recombination (right).

**Sex Chromosomes**

The emergence of isogamy, ie, mating types, was not considered in the explanation for the evolution of two-step meiosis with recombination above. However, the presence of allosomes - XY in animals and ZW in birds, some fish, reptiles, insects, etc. – can also be explained as a mechanism by which a haploid genome may vary the amount of learning it experiences when paired with another to form a diploid organism. Importantly, taking the view of the constituent haploid genomes, the presence of an heterogametic sex creates the situation where, as evolution converges upon optima, a given haploid containing the common (X or Z) autosome will typically experience two different fitness values simultaneously within a population due to genetic differences between the two sexes; two fitness contributions from the common autosome will almost always exist with two mating types. It is proposed that the extra (approximate) fitness value information can prove beneficial to the learning/generalisation process described above by adding further landscape smoothing.

To introduce autosomes and allosomes to the above diploid model, the original pair of haploid genomes of length $N$ are each subdivided into $n=2$ equally sized chromosomes, ie, there are $2n$ chromosomes per diploid. A (converged) sub-population of a homogametic sex is said to exist along with a (converged) sub-population of a heterogametic sex. No functional differentiation is imposed upon the heterogametic sex fitness function here; the fitness landscapes of both sexes are identical.

Autosomes undergo two-step meiosis with recombination, as above, whereas allosomes do not undergo recombination. The sex of the offspring is determined by which allosome is (randomly) selected from the heterogametic sex. Once the resulting overall diploid genome is created, mutation is applied to each haploid as before (Figure 6). The fitness contribution of the haploid genomes is their average, as above. For example, when X-inactivation occurs in mammals the choice is typically random per cell lineage in the placenta and hence the fitness contribution of the allosomes remains a composite of the two chromosomes.

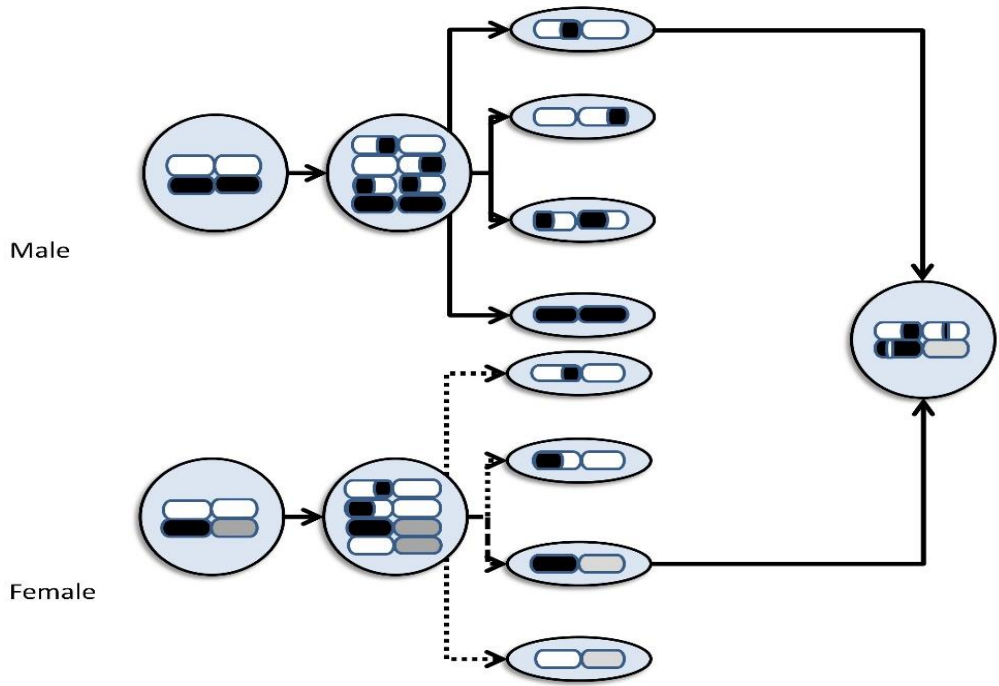

Figure 6. Showing the haploid-diploid cycle with two-step meiosis as implemented with converged sub-populations of females and males (ZW system assumed).

Figure 7 shows examples of how the benefits of sex chromosomes can vary with landscape ruggedness and size. Note the average fitness of the heterogametic and homogametic sexes is shown. An equivalent (converged) population of hermaphrodites is used for comparison here, where recombination is not used for the second chromosome. As can be seen, for $N$=20, two sexes prove beneficial for $K$>5 (T-test, $p$<0.05) over the hermaphrodite. In contrast, with $N$=100, two sexes prove detrimental under the same conditions (T-test, $p$<0.05). Therefore the improvement in fitness seen for sex cannot simply be attributed to the presence of two explicit sub-populations. There is no significant difference in fitness seen for $K$<6 (T-test, $p$≥0.05) for either $N$. It can also be noted that including recombination in the second chromosomes (allosomes) of the hermaphrodites does not significantly affect their fitness under any conditions explored (not shown, T-test, $p$≥0.05). Not allowing recombination between the autosomes in the two sexes has no significant effect when $N$=100 (not shown, T-test, $p$≥0.05), as is the case with $N$=20 until $K$>2 where its removal is detrimental (not shown, T-test, $p$<0.05).

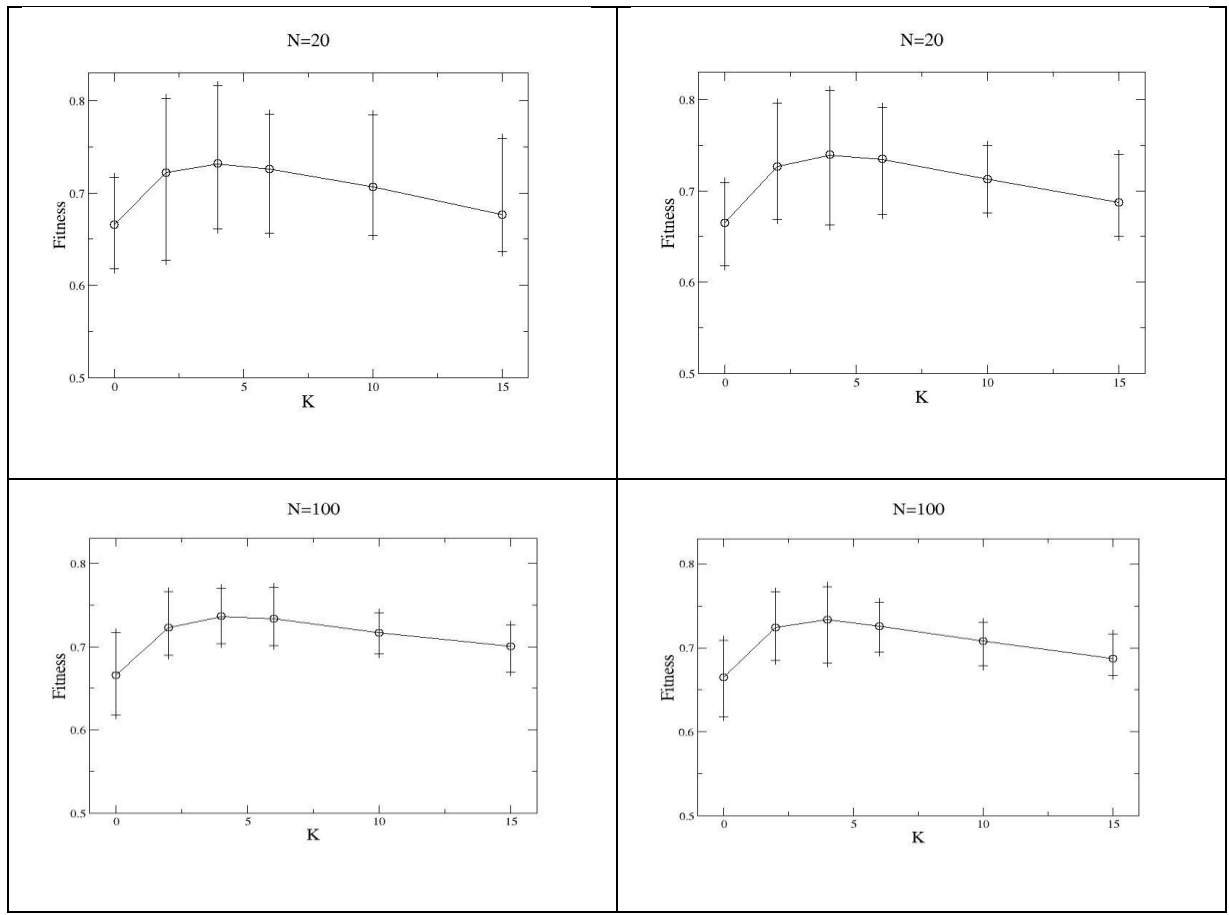

Figure 7. Showing typical behaviour and fitness reached after 20,000 generations on landscapes of varying ruggedness ($K$) and size ($N$) for hermaphrodite diploids (left) or with two sexes (right).

Following [6], it was shown in [7] that increasing the amount of learning under a simple scheme with haploid genomes gives increasing benefit for $K>6$ with $N=20$, whereas increasing the amount of learning decreases fitness for $K>4$ with $N=100$. As noted above, it is here suggested that the presence of sex chromosomes creates another mechanism through which the fitness landscape of the constituent haploids with the common autosome is potentially smoothed. That is, as the degree of heterogeneity between the two haploids in the homogenetic sex converges (Figure 5, right) two different fitness values are maintained due to the fitness obtained within the heterogametic sex. This is in contrast to hermaphrodites where eventual near convergence of the two haploids reduces the amount of learning/smoothing as an optimum is found. Therefore the results in [7] predict those seen here with sex chromosomes added almost exactly: the extra learning mechanism is useful for higher $K$ when $N=20$ and detrimental for higher $K$ when $N=100$. Although the extra smoothing proves beneficial on less rugged landscapes (*K*>4 vs *K*>6) than in the standard case with *N*=20 perhaps due to the convergence of the two haploids over time reducing the amount of learning experienced from their partnering.

It can be noted that, whilst varying between reproducing with a member of the opposite sex and as a hermaphrodite is seemingly relatively common in nature (eg, see [3]), no significant change in behaviour was seen here for a variety of ratios for the parameters explored (not shown). However, the explanation presented that either form of reproduction provides a difference in the amount of learning exhibited suggests a selective advantage for species able to suitably tune the ratio between the two based upon current conditions.

**Allosomes: Breaking Symmetry**

The missing allosome (Z0/X0) form of isogamy also enables a variation in the amount of learning since one mating type will provide the "pure" fitness of a single allosome. A general benefit of two types of allosome can be identified when the diploid fitness landscape is considered. As shown in Figure 5 (right), hermaphrodites exploiting two-step meiosis with recombination can be expected to eventually converge upon organisms carrying two copies of the same (or very nearly) haploid genome. This means that over evolutionary time they become increasingly restricted to the region of symmetry within the diploid fitness landscape, ie, the region where constituent haploid genome A is genetically similar to haploid genome B. However, high levels of fitness may exist in other areas of the diploid landscape, particularly with two mating types, as shown in Figure 8. One way to avoid a near complete set of homozygotes is to maintain a region(s) in the haploid genome where recombination does not occur, thereby inhibiting the assimilation process. It is here suggested this is what two types of allosome enables over a single type. Asexual reproduction and one-step meiosis, ie, a haploid-diploid cycle without recombination, can also hinder convergence (Figure 5, left). Whilst the results above suggest two-step meiosis with recombination is generally beneficial over asexuality and one-step meiosis due to the increased amount of learning, there is some overlap in fitness reached at lower levels of landscape ruggedness. For example, Parabasalid are known to exploit simple syngamy and since they have lost their mitochondria may be seen to exist on less rugged landscapes (after [8]). It can be speculated that they may therefore be exploiting the potential to move away from the region of symmetry within their diploid fitness landscapes. More generally, having adopted the more beneficial two-step meiosis with recombination, two types of allosome enables eukaryotes to maintain heterozygotes for some regions of their overall genome space in a relatively controlled manner. This becomes particularly significant if the presence of mating types also causes an increase in fitness in the region of symmetry, eg, through dimorphism, as will be discussed below.

Figure 9 shows results from introducing heterogeneity/difference into the fitness landscape of the two mating types used above. Here the fitness values in the table for the Z chromosome are constructed as usual and those for the W chromosome are made by subtracting the corresponding value from 1.0, ie, so that the opposite gene values are preferred. As expected, performance is typically improved when recombination between Z and W chromosomes is not allowed but is between Z chromosomes ($K$>4, T-test $p$<0.05) when $N$=100. Results with $N$=20 do not show any significant difference in fitnesses reached (not shown, T-test $p$≥0.05). The equivalent hermaphrodite is always worse, regardless of $N$ and $K$ (not shown, T-test $p$<0.05).

Hence this explains such things as why recombination does not typically occur between allosome types but why recombination over some regions is sometimes seen; the degree of difference is controllable. Relatedly, the two sex chromosomes can be of different sizes which is also potentially correlated to the degree of heterogeneity required between the two haploids to reach the higher fitness areas. Note the situation can be reversed from the ZW system in the XY system since which mating type maintains the single copy of the allosome is not important in exploiting the benefits described. It also helps to explain why more than two mating types exist in some species - the presence of more than one high fitness region away from the area of symmetry in the fitness landscape is exploitable by the maintenance of a corresponding mating type per region. The aforementioned case of higher optima existing in the region of symmetry for the homogametic sex through the existence of two mating types over hermaphrodites would explain the maintenance of sex in even the most unchanging environments. Conversely, that the existence of such regions may vary temporally could help to explain environmental sex determination mechanisms. Moreover, similar reasoning would seem to apply regardless of the details of the mechanism, eg, for polygenic sex, cytoplasmic control, etc.

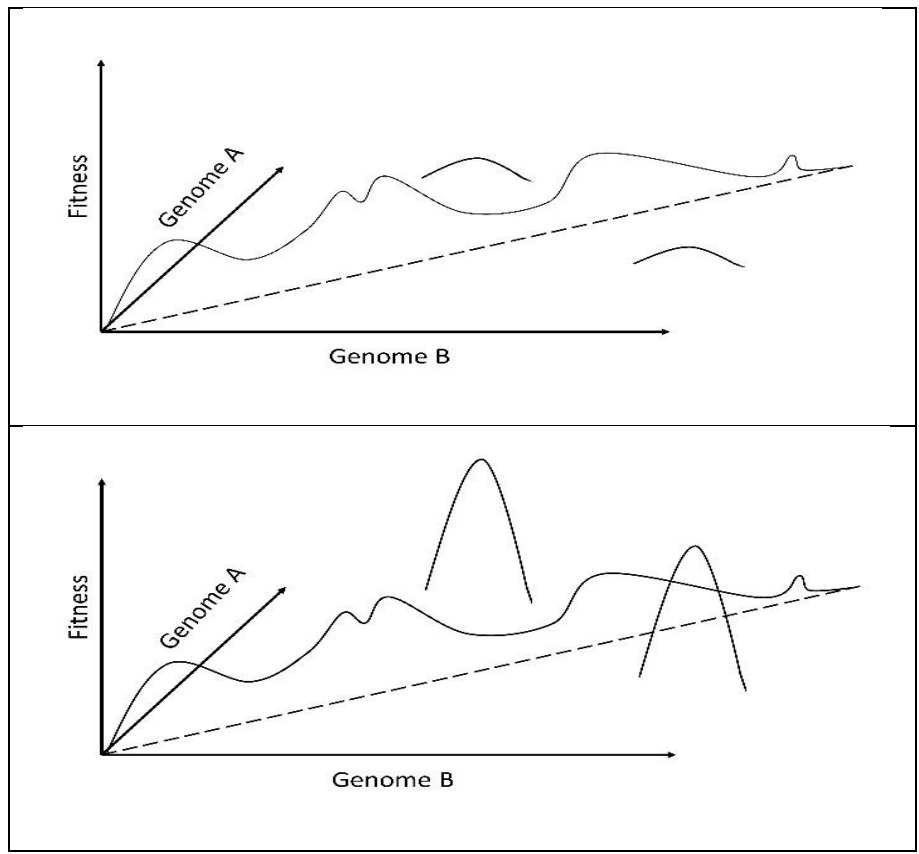

Figure 8. Showing simple example fitness landscapes where mating types are not (top) and are expected (bottom) to prove beneficial. In the second case, high optima also exist away from the area of genome symmetry within the diploid landscape. The existence of sex chromosomes enables the males (females) to maintain genomes corresponding to the end-point on the axis of possible gene combinations and therefore occupy the outlier optima.

Perhaps most significantly, once allosomes have evolved and two (or more) mating types become established, a coevolutionary scenario emerges between the two sexes. That is, the fitness landscapes of each sex will typically be both different and moving due to changes in the opposite sex. The emergence of intersexual selection as a consequence is now explored using the coevolutionary version of the NK model – the NKCS model.

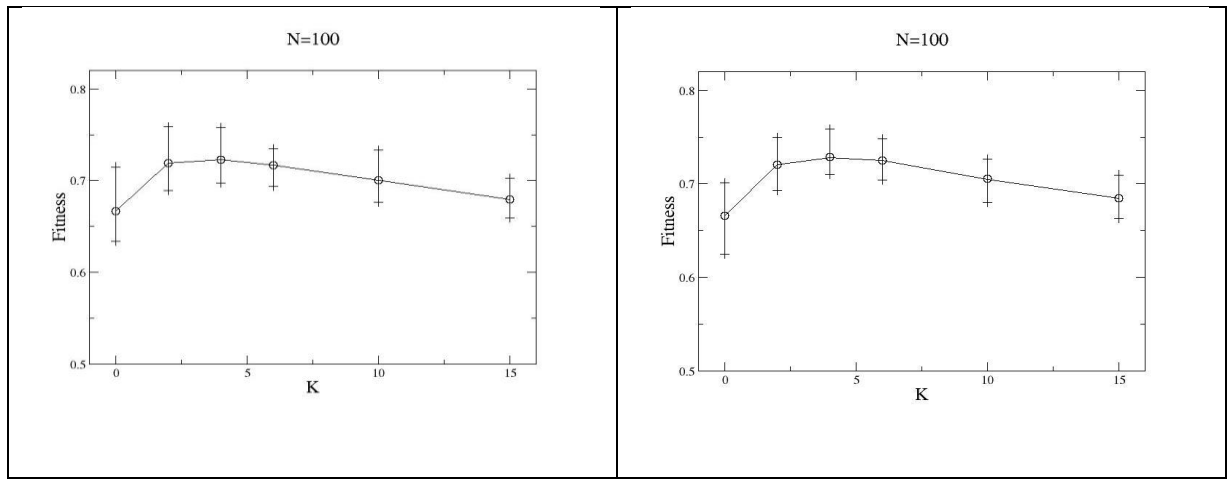

Figure 9. Showing typical behaviour and fitness reached after 20,000 generations on landscapes of varying ruggedness (*K*) with a heterogeneous component in the sex chromosome, with (left) and without (right) recombination between X and Y chromosomes.

**The NKCS Model: Asexual Coevolution**

Kauffman and Johnsen [15] introduced the NKCS model to enable the study of various aspects of *co*evolution. Extending the NK model, each gene is said to also depend upon *C* randomly chosen traits in each of the other *S* species with which it interacts. Altering *C*, with respect to *N*, changes how dramatically adaptive moves by each species deform the landscape(s) of its partner(s), where increasing *C* typically increases the time to equilibrium. Again, for each of the possible $K+(SxC)$ interactions, a table of $2^{(K+(SxC)+1)}$ fitnesses is created for each gene, with all entries in the range 0.0 to 1.0, such that there is one fitness for each combination of traits. Such tables are created for each species (Figure 10). Again, a population is of size one and a species evolves by making a random change to one randomly chosen gene per generation.

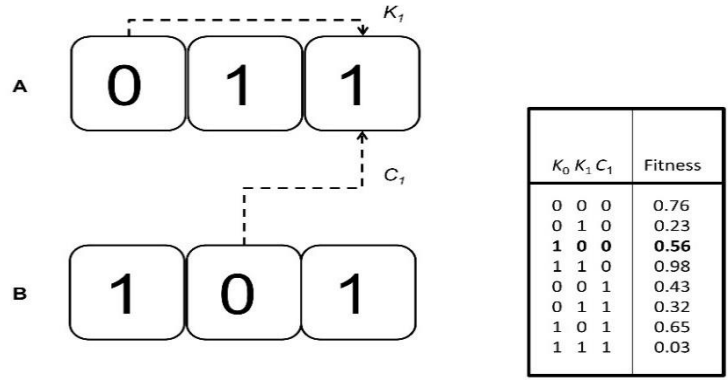

Figure 10. The NKCS model: Each gene is connected to *K* randomly chosen local genes and to *C* randomly chosen genes in each of the *S* other species. Connections and table shown for one gene in one species for clarity. Here *N*=3, *K*=1, *C*=1, *S*=1.

Figure 11 shows example behaviour for one of two coevolving species where the parameters of each are the same and hence behaviour is symmetrical. The effects of mutual fitness landscape movement are clearly seen. All results reported in this paper are the average of 10 runs (random start points) on each of 10 NKCS functions, ie, 100 runs, for 20,000 generations. Here 0≤*K*≤10, 1≤*C*≤7, for *N*=20 and *N*=100. Figure 12 shows how increasing the degree of connectedness (*C*) between the two landscapes causes fitness levels to fall significantly (T-test, $p<0.05$) when *C*≥*K* for *N*=20. Note this change in behaviour around *C*=*K* was suggested as significant in [13], where *N*=24 was used throughout. However, Figure 12 also shows how with *N*=100 fitness *always* falls significantly with increasing *C* (T-test, $p<0.05$), regardless of *K*.

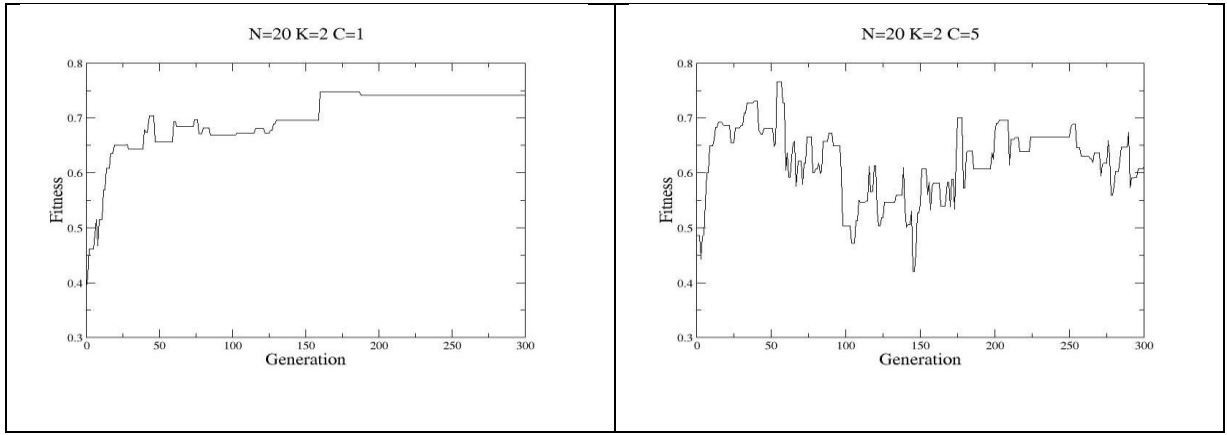

Figure 11. Showing example single runs of the typical behaviour of the standard NKCS model of coevolution with different degrees of landscape coupling ($C$).

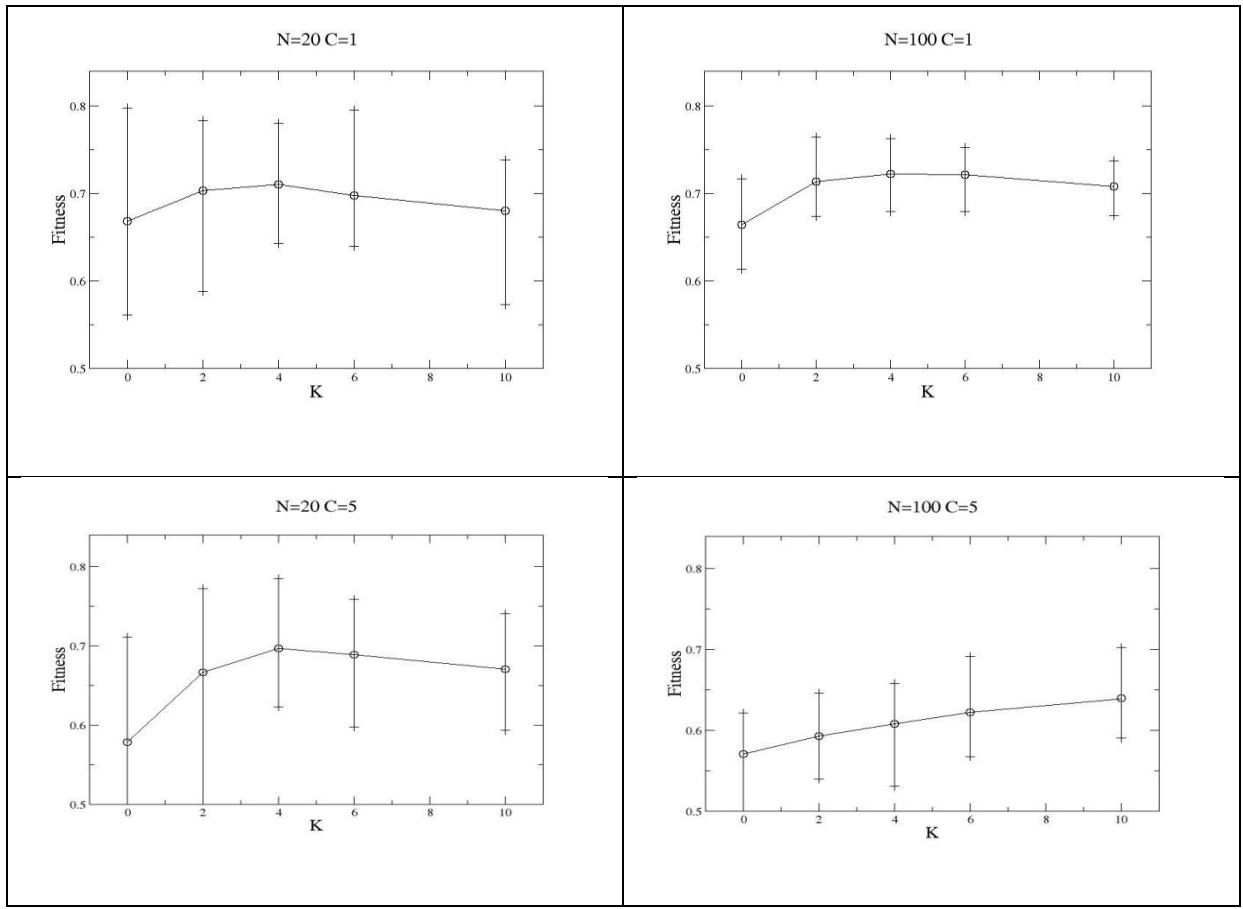

Figure 12. Showing the fitness reached by coevolving asexual haploids after 20,000 generations on landscapes of varying ruggedness ($K$), coupling ($C$), and length ($N$). Error bars show min and max values.

## Mating Types in the NKCS Model: Coevolving Sexes

Figure 13 shows how epistatic connections are considered in a diploid where each gene depends on $K$ local genes and $C$ genes in each of the genomes of its partner. Here each gene can have two different fitness values depending upon the homogeneity of its partner. Hence the overall fitness total of a diploid is in the range $[0, 4N]$ and so the total is normalised by $4N$ to determine its fitness for selection, applied as above.

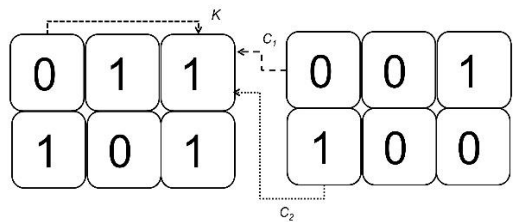

Figure 13. The NKCS model extended to diploids: Each gene is connected to $K$ randomly chosen local genes and to $C$ randomly chosen genes in *each* of the genomes in the other species/populations. Connections for one gene in one genome of one species/population shown for clarity. The fitness contribution of each $K$ and $C$ combination is found from the associated fitness table. Total fitnesses are normalised by $4N$.

Figure 14 shows the typical behaviour seen for various combinations of genome size ($N$) and within ($K$) and between ($C$) sex dependence. As anticipated by the traditional model above, increasing the dependence between the sexes increases the effects of their adaptations upon each other, with the average of the male and female fitnesses dropping as a consequence.

Whilst direct comparison to the traditional model is not valid, comparison can be made to two asexual diploid populations coevolved on the same set of fitness landscapes. That is, the performance of sex in a coevolutionary context can be explored here. Figure 15 shows examples of how the same general benefit as above (Figure 4) is seen with $N=20$, regardless of $C$, with sex resulting in higher fitness (T-test, $p<0.05$) when $K>2$. The same is seen when $N=100$, although only for $K>6$ when $C=1$.

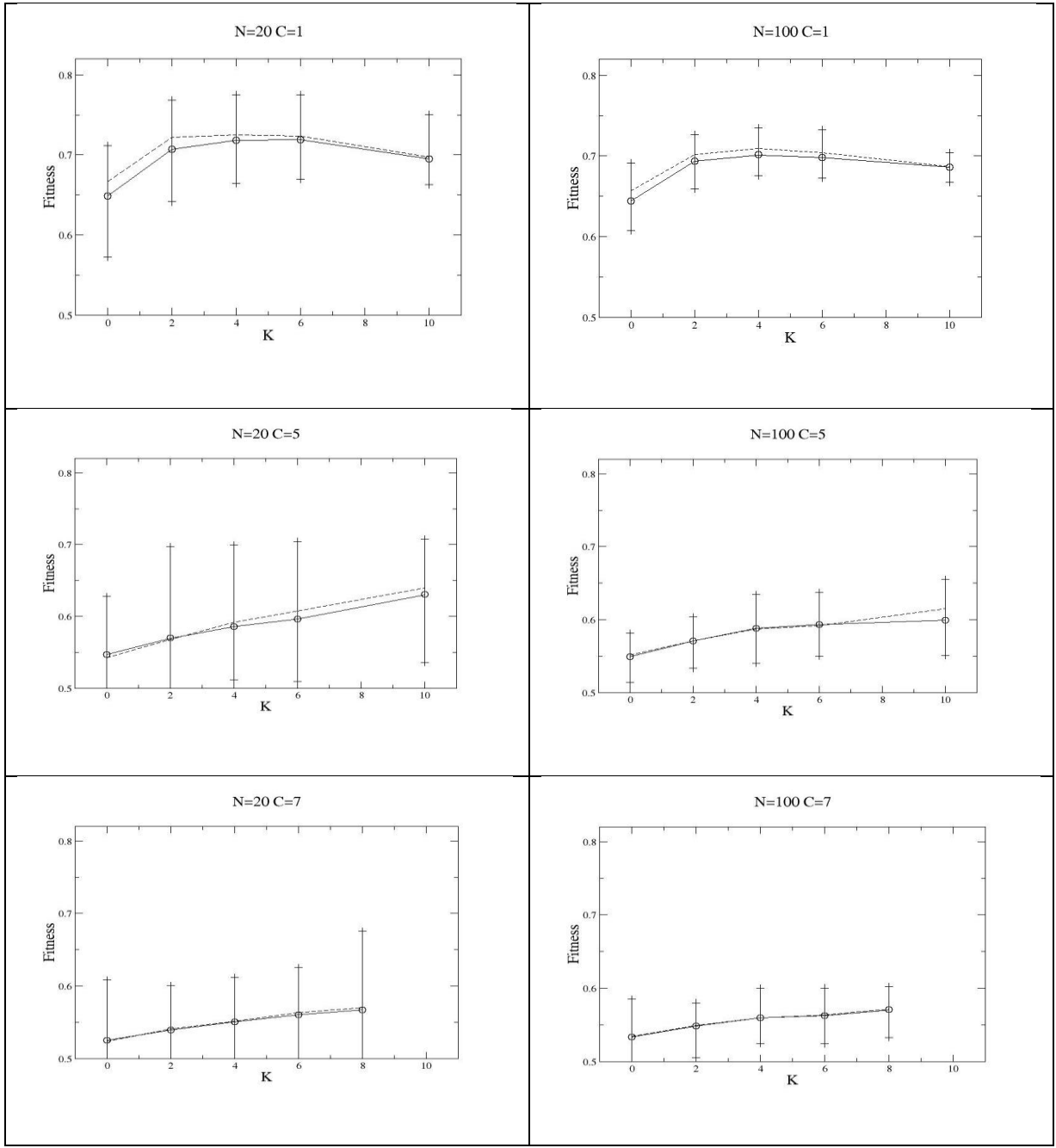

Figure 14. Showing the fitness reached after 20,000 generations for a sexual diploid species treated as a coevolutionary scenario on landscapes of varying ruggedness (*K*), coupling (*C*), and length (*N*). The fitness shown is the average of the males and females. The dashed line shows female fitness.

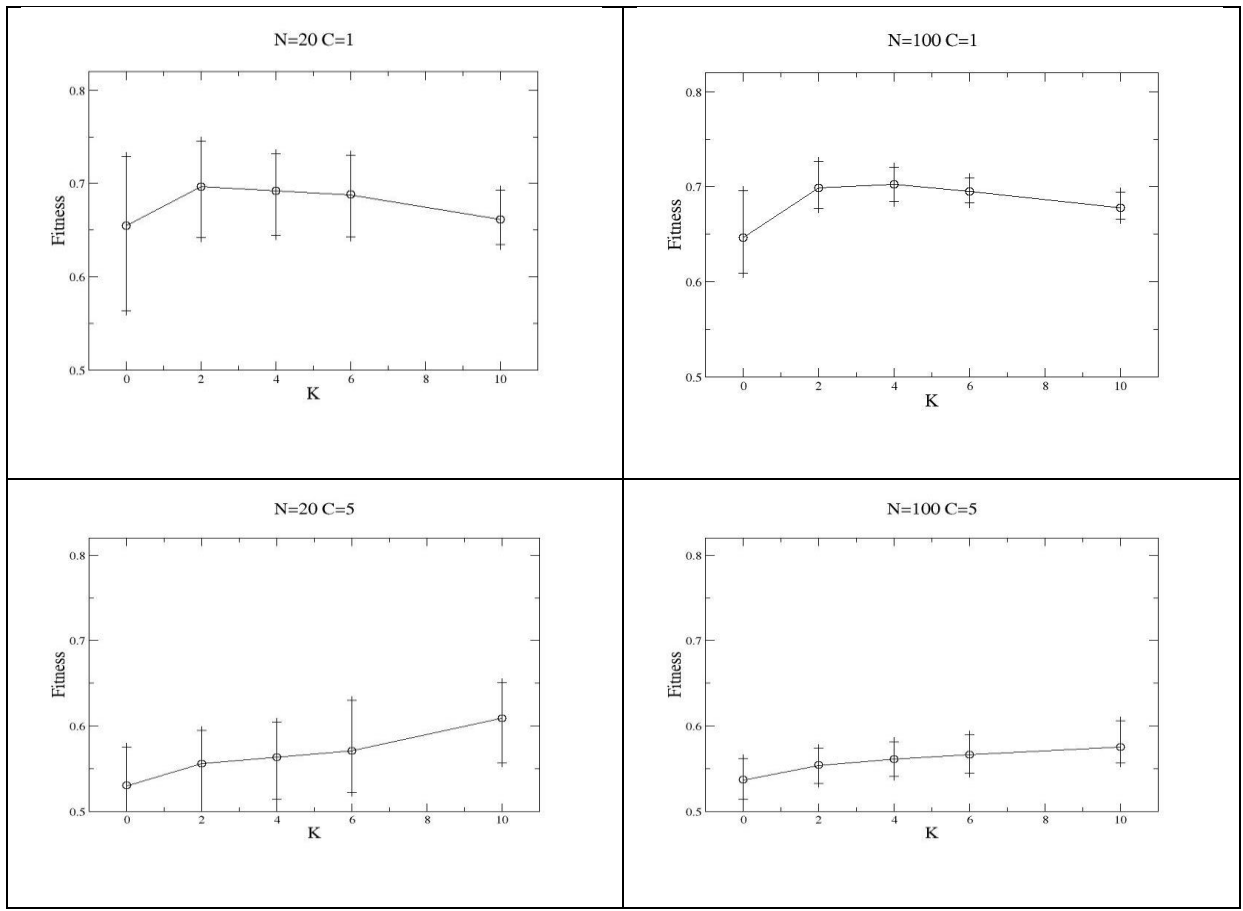

Figure 15. Showing the fitness reached by two populations of coevolving asexual diploids after 20,000 generations on landscapes of varying ruggedness ($K$), coupling ($C$), and length ($N$). Error bars show min and max values.

**Sexual Selection in the NKCS Model**

Intersexual selection – whether pre and/or post mating – can be viewed as the imposition of one or more preferred traits by one sex upon the other. This selective pressure is a component of the overall selective pressure an organism experiences: a male peacock's train is no more an extravagance or handicap than a giraffe's neck - both are simply the result of selection under coevolution. It can also be noted that, as with any trait, any correlation between the imposed trait(s) and any other element of an organism's selective pressure will typically be coincidental. And correlations, or a lack therefore, are likely to change with adaptations made by other organisms within the given organism's coevolutionary environment over time. That is, "good gene" selection can be expected (eg, [9]) but not ubiquitously (eg, [1]).

Under this view, the above NKCS model can be altered to include mate choice by the addition of an extra binary string said to represent a preferred set of *N* traits, created at random with the rest of the fitness function, by the choosing sex. Fitness for selection then becomes the fitness used in the previous section and the fraction of genes which match the imposed traits in both genomes of the other sex. Since females are typically the limiting factor in reproduction due to the frequency of opportunity, their investment in offspring rearing, etc., the females (ZW) are assigned the random binary string of required traits here. That is, overall fitness for selection in the male (ZZ) population is now in the range [0.0, 2.0] before normalisation.

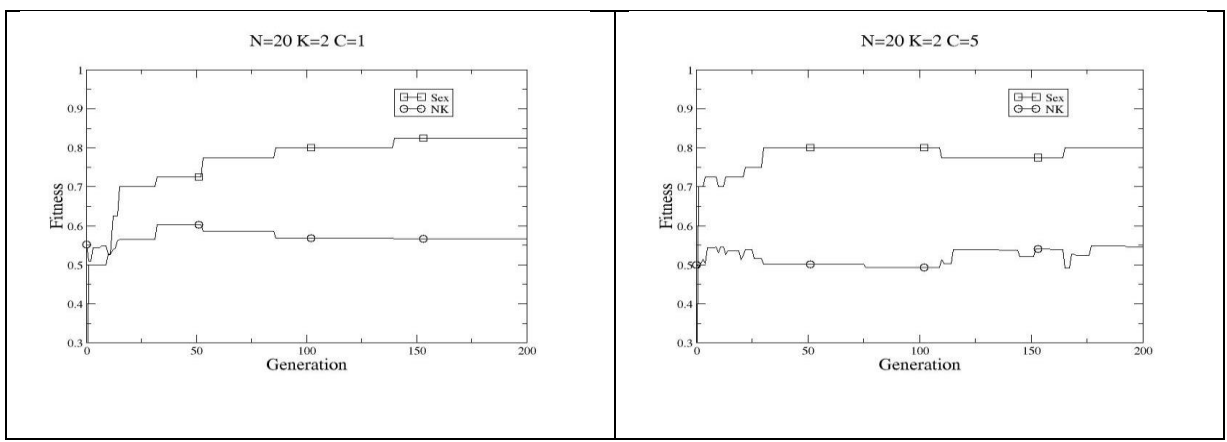

Figure 16. Showing example single runs of the typical behaviour of the NKCS model of coevolution between two sexes with sexual selection and different degrees of landscape coupling (*C*). Contrast with Figure 11.

Figure 16 shows example (co)evolutionary dynamics for the average species where the males' fitness includes the degree of match to the females' trait preferences. As can be seen, in contrast to the typical dynamics seen in such coevolutionary models above (Figure 11), the effects of mutual fitness landscape movement can be reduced as *C* increases. Comparing Figure 14 to Figure 17 shows how for *N*=20, with *C*<5, fitnesses – whether the females' or the species' average - are worse than without the intersexual selection, regardless of *K* (T-test, $p<0.05$). When *C*=5 the simple mechanism proves beneficial when *K*<4 and when *C*=7 for *K*<8 (T-test, $p<0.05$). In contrast, no benefit is seen when *N*=100 for all *K* (T-test, $p\geq 0.05$). Figure 17 also shows how the sexual selection fitness component of the males' fitnesses generally decreases with increasing *K* and is almost always lower when *N*=100 compared to when *N*=20.

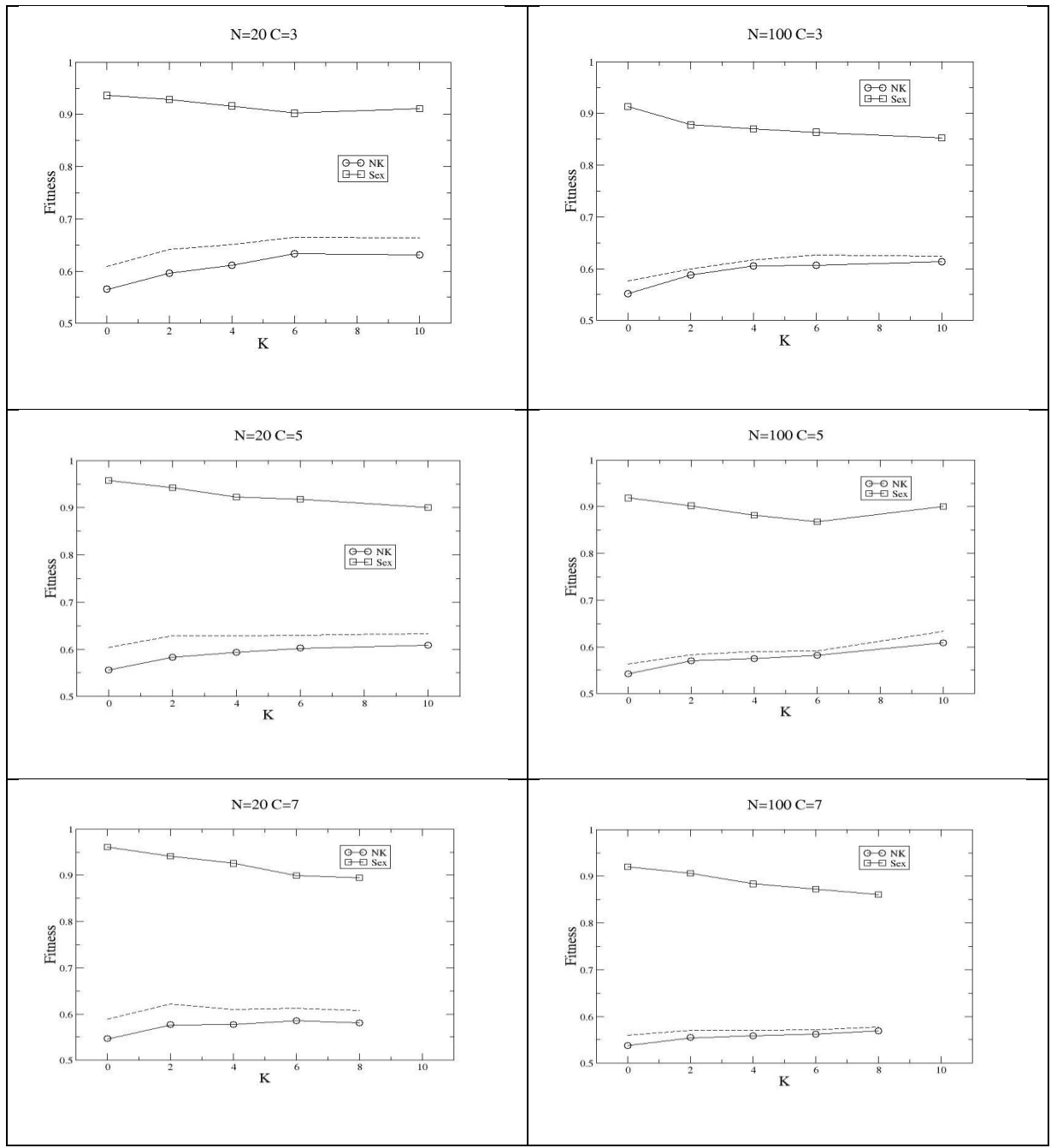

Figure 17. Showing the fitness (original and sexual) reached after 20,000 generations for a diploid species treated as a coevolutionary scenario exploiting simple sexual selection by females on landscapes of varying ruggedness (*K*), coupling (*C*), and length (*N*). The fitnesses shown are the average of the males and females. Dashed lines show female fitness. Error bars are not shown for clarity.

The same general results are seen if the sexual selection component of male fitness is calculated using the fraction of trait match on the first haploid genome only rather than both, and is significantly worse if a single matching genome is chosen at random per generation (not shown). Moreover, if the template is reduced in size, ie, the number of traits is reduced, from $N$ to $N/2$ such that matching only occurs on the two sex chromosomes, results remain the same (not shown).

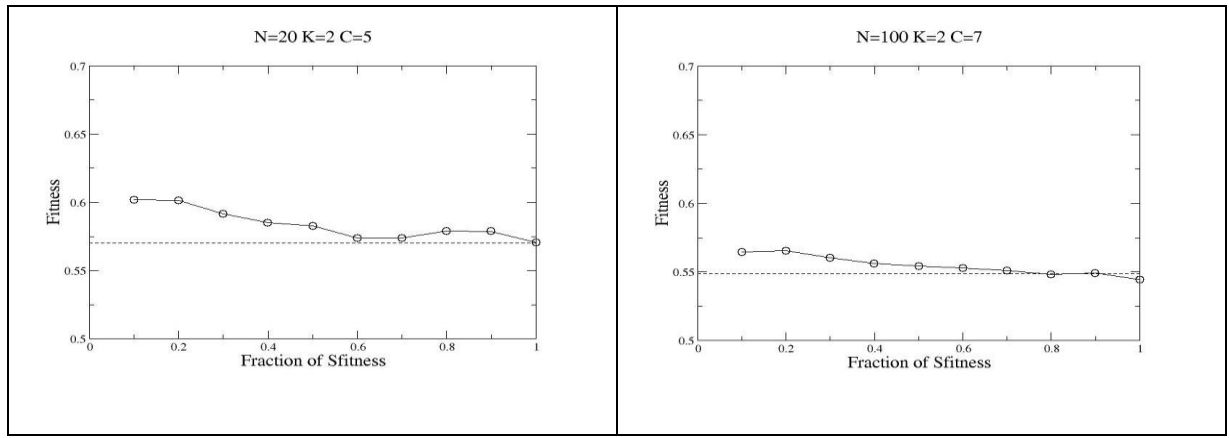

Figure 18. Showing examples of how varying the weighting/strength of sexual selection affects species' fitness after 20,000 generations. The dashed line shows fitness without sexual selection (Figure 14).

In the above, intersexual selection fitness was equally weighted (0.5) with the underlying NKCS fitness function: male selection fitness was the sum of the two values. Figure 18 shows examples of how varying the weighting can vary the benefits of sexual selection, particularly for lower values (<0.5). These results suggest that tuning the strength of sexual selection to match the underlying coevolutionary dynamics represents an important adaptive mechanism for a species (eg, see [11] for discussion in dynamic environments).

Thus far the preferred traits of the choosing sex were incorporated as a separate component of the overall fitness landscape of the other sex – and they were unchanged throughout evolution. Of course, this unlikely to be the case generally. Figure 19 shows examples of how varying the frequency of changing one randomly chosen preferred trait can vary the benefits of intersexual selection. As can be seen, the benefit of the simple mechanism is lost as the frequency of change increases and so they can be expected to evolve relatively slowly. It can be noted that in birds, whilst sexual selection is known to accelerate the evolution of mating traits compared to other traits in males, no similar effect is seen in females [19].

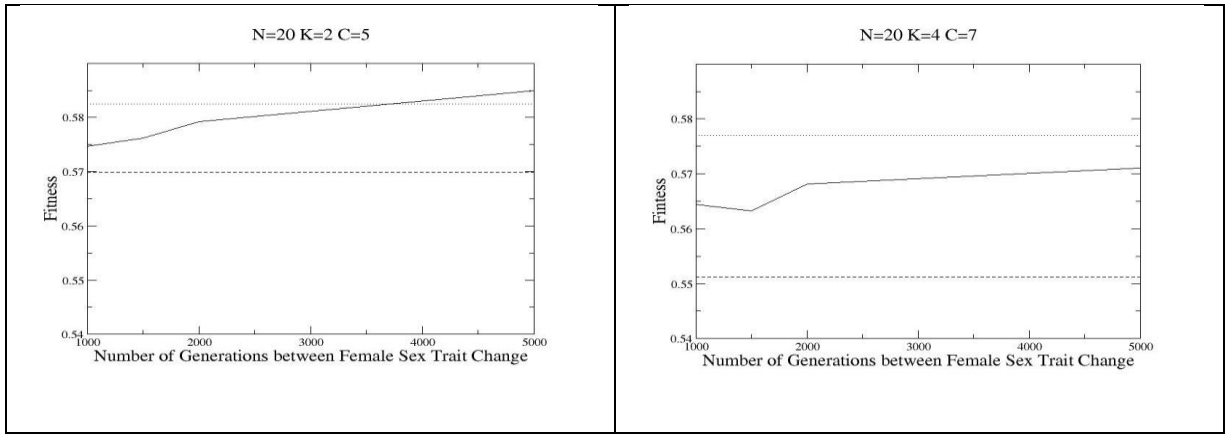

Figure 19. Showing examples of how the rate of varying the preferred traits of the female affects species' fitness after 20,000 generations. The dotted line shows fitness with unchanging traits (Figure 17) and the dashed line shows fitness without sexual selection (Figure 14).

Finally, Figure 20 shows results from when both males and females are exploiting the simple sexual selection mechanism. When $N$=20 and $C$<5, fitnesses are worse than without sexual selection for all $K$ (T-test, $p$<0.05), the same when $C$=5 for all $K$ (T-test, $p$≥0.05), and when $C$=7 fitnesses are improved for $K$>2 (T-test, $p$<0.05). When $N$=100 and $C$<5 fitnesses are worse for all $K$ (T-test, $p$<0.05), when $C$=5 they are worse when $K$<2 and better when $K$>4, and when $C$=7 fitnesses are improved when $K$>2 (T-test, $p$<0.05). Recall that no benefit was seen from one mate exploiting sexual selection in the more complex case of $N$=100 above. Further, in comparison to one mate using the mechanism, there is no significant change when $N$=20. When $N$=100, for $C$<7 and $K$<4 fitnesses are worse but improved when $K$>4 (T-test, $p$<0.05). Similarly, when $C$=7 fitness are worse when $K$<2 and improved when $K$>2. Hence a benefit is again typically seen with increasing $C$.

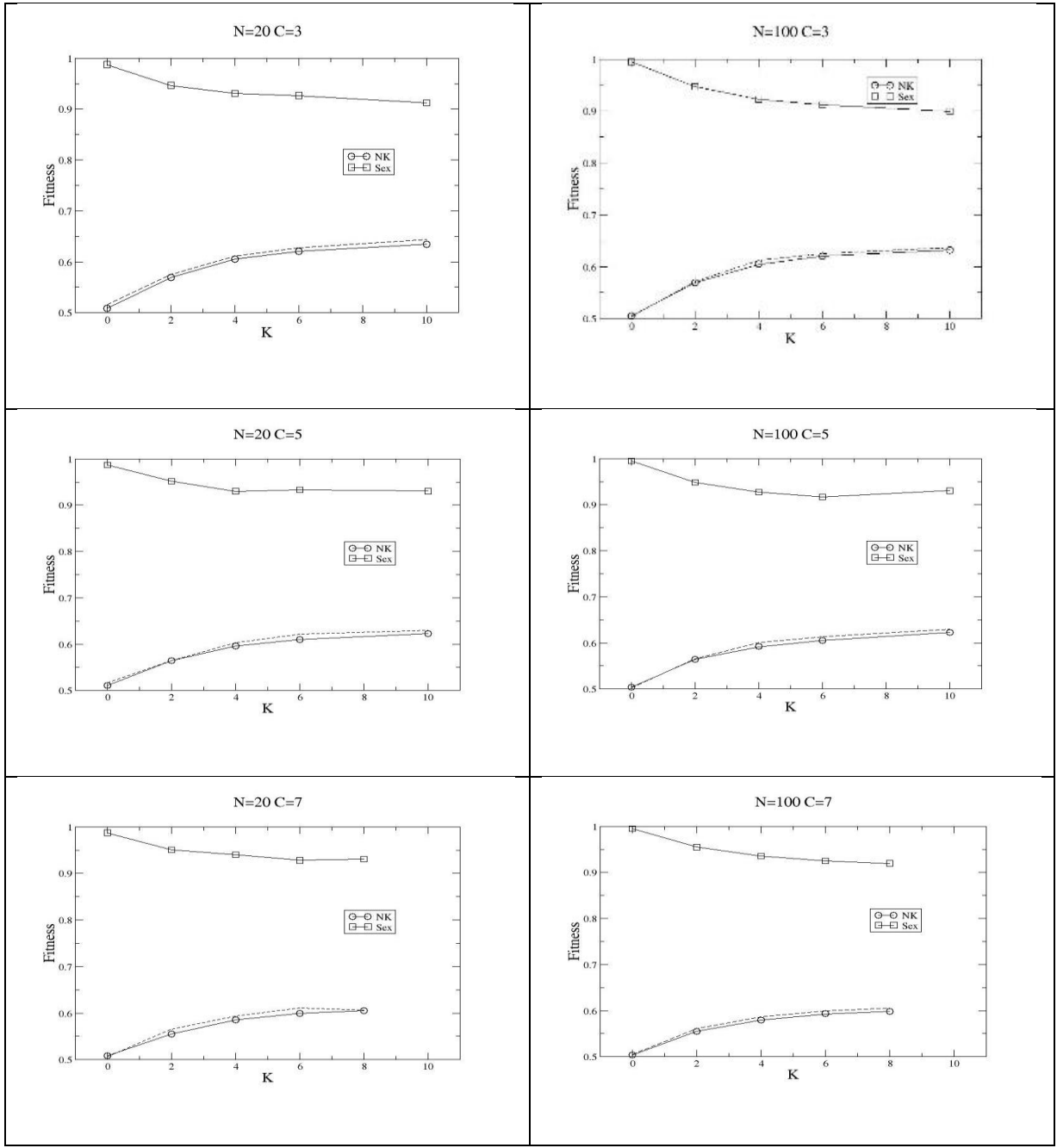

Figure 20. Showing the fitness (original and sexual) reached after 20,000 generations for a sexual diploid species treated as a coevolutionary scenario exploiting simple sexual selection by *both* females and males on landscapes of varying ruggedness (*K*), coupling (*C*), and length (*N*). The fitnesses shown are the averages of the males and females. Dashed lines show female original fitness. Error bars are not shown for clarity.

**Conclusion**

This paper has considered intersexual selection as within species coevolution caused by the emergence of sex chromosomes. The evolution of allosomes was explored and a beneficial fitness landscape smoothing effect from the extra (approximate) fitness value information was found for smaller, more rugged landscapes. Moreover, the same reasoning applies equally well to the emergence of a single allosome, although the effect from the extra fitness value would potentially reduce over time as the homogametic sex converges depending upon any dosage effects, etc. It can also be noted that X inactivation of the paternal/maternal chromosome in all cells, eg, as in marsupials, as opposed the random scheme used here, eg, as in humans, represents another mechanism through which to vary the amount of learning experienced.

Using the NKCS model, the resulting coevolutionary behaviour between two mating types have been explored with and without an extra intersexual selection component. As anticipated by the traditional model, increasing the dependence between the sexes increases the effects of their adaptations upon each other, with fitnesses dropping as a consequence. Possible sources of high dependence between the sexes include the obligate nature of their reproduction, offspring rearing, social structures, etc.

It has been suggested here that in its simplest form, mate choice can be viewed as the imposition of one or more preferred traits by one sex upon the other. Under the coevolutionary view, this represents a mechanism by which the degrees of evolutionary freedom of one sex are reduced by the other, ie, it reduces the amount of fitness landscape movement if the preferred traits represent a relatively steady (target) component of the overall selection pressure experienced. This might emerge since females consistently preferring a certain trait(s) in their male partner during their lifetime would potentially reduce the variance they experience in nesting, offspring rearing, etc. in comparison to selecting a mate at random each time. Reducing variance over evolutionary time typically increases fitness. Moreover, this is seen to be true for both female fitness and the overall species/population fitness (see [10] for related discussions).

**References**

1. Achorn, A. & Rosenthal, G. (2020). It's not about him: mismeasuring 'good genes' in sexual selection. *Trends in Ecology and Evolution*, 35, 206-219.

2. Alonzo, S.H. & Servedio, M.R (2019). Grey zones of sexual selection: why is finding a modern definition so hard? *Proceedings of the Royal Society B,* 286, 20191325

3. Bachtrog, D. et al. (2104). Sex determination: Why so many ways of doing it? *PLOS Biology,* 12, e10001899

4. Baldwin, J. M. (1896). A new factor in evolution. *American Naturalist,* 30, 441–451.

5. Bernstein, H. & Bernstein, C. (2010). Evolutionary origin of recombination during meiosis. *BioScience*, 60, 498-505

6. Bull, L. (1999). On the Baldwin effect. *Artificial Life,* 5, 241–246

7. Bull, L. (2017). The evolution of sex through the Baldwin effect. *Artificial Life,* 23, 481-492

8. Bull, L. & Fogarty, T.C. (1996). Artificial symbiogenesis. *Artificial Life,* 2, 269-292.

9. Byers, J. & Waits, L. (2006). Good genes sexual selection in nature. *Proceedings of the National Academy of Sciences*, 103, 16343-16345.

10. Cally, J.R., Stuart-Fox, D. & Holman, L. (2019). Meta-analytic evidence that sexual selection improves population fitness. *Nature Communications*, 10, 2017.

11. Candolin, U. & Heuschele, J. (2008). Is sexual selection beneficial during adaptation to environmental change? *Trends in Ecology and Evolution*, 23, 446-452.

12. Hinton, G. E., & Nowlan, S. J. (1987). How learning can guide evolution. *Complex Systems,* 1, 495–502.

13. Kauffman, S. A. (1993). *The Origins of Order: Self-organisation and Selection in Evolution*. New York, NY: Oxford University Press.

14. Kauffman, S. A., & Levin, S. (1987). Towards a general theory of adaptive walks on rugged landscapes. J*ournal of Theoretical Biology*, 128, 11–45

15. Kauffman, S. A., & Johnsen, S. (1992). Co-evolution to the edge of chaos: Coupled fitness landscapes, poised states and co-evolutionary avalanches. In C. G. Langton, C. Taylor, J. D. Farmer, & S. Rasmussen (Eds.), *Artificial Life II* (pp. 325–370). Reading, MA: Addison-Wesley.

16. Maynard Smith, J. & Szathmary, E. (1995). *The Major Transitions in Evolution.* WH Freeman, Oxford

17. Orr, H. & Otto, S. (1994) Does diploidy increase the rate of adaptation? *Genetics*, 136, 1475-1480.

18. Rowe, L. & Arnqvist, G. (2002). Sexually antagonistic coevolution in a mating system: Combining experimental and comparative approaches to address evolutionary processes. *International Journal of Organic Evolution,* 56, 754–67.

19. Seddon, N., Boterro, C., Tobias, J.A., Dunn, P., MacGregor, H.E.A., Rubenstein, D., Uy, A., Weir, J.T., Whittingham, L. & Safari, R. (2013). Sexual selection accelerates signal evolution during speciation in birds. *Proceedings of the Royal Society B*, 280, 20131065.